\documentclass{WileyMSP-template}
\usepackage{xcolor}
\usepackage{graphicx}
\usepackage{amsmath} 
\usepackage{amsfonts}
\usepackage{amssymb}
\usepackage[normalem]{ulem}
\usepackage{float} 

\begin{document}

\title{Plasmonic Metasurfaces with Structural Chirality and Pseudo-Chirality for Enhanced Circular Dichroism and Enantiomeric Recognition}

\maketitle


\author{Giovanna Palermo*}
\author{Bryan Guilcapi}
\author{Radoslaw Kolkowski}
\author{Alexa Guglielmelli}
\author{Dante M. Aceti}
\author{Liliana Valente}
\author{Joseph Zyss}
\author{Lucia Petti*}
\author{Giuseppe Strangi*}


\begin{affiliations}
G. Palermo, A. Guglielmelli, D. M. Aceti, L. Valente\\
Department of Physics, NLHT Lab - University of Calabria and CNR-NANOTEC Istituto di Nanotecnologia, 87036 - Rende, Italy\\
Email Address: giovanna.palermo@unical.it

R. Kolkowski\\
Department of Applied Physics, Aalto University, P.O.Box 13500, Aalto FI-00076, Finland

J. Zyss\\
Laboratoire Lumière, Matière et Interfaces (LuMin), ENS Paris-Saclay, CentraleSupélec,
CNRS, Université Paris-Saclay, 91190 Gif-sur-Yvette, France

L. Petti, B. Guilcapi\\
Institute of Applied Sciences and Intelligent Systems ”E. Caianiello” CNR, 80078 Pozzuoli, Italy\\
Email Address:lucia.petti@isasi.cnr.it

G. Strangi\\
Department of Physics, Case Western Reserve University, 2076 Adelbert Rd, Cleveland, Ohio 44106, USA\\
Department of Physics, NLHT Lab - University of Calabria and CNR-NANOTEC Istituto di Nanotecnologia, 87036 - Rende, Italy\\
Email Address: giuseppe.strangi@fis.unical.it

\end{affiliations}




\justify

\begin{abstract}
We present the design and optical characterization of a plasmonic metasurface engineered to exhibit strong polarization anisotropy under both linearly and circularly polarized light. The metasurface consists of geometrically asymmetric gold nanostructures arranged periodically on a glass substrate. Each nanostructure is formed by the fusion of three equilateral triangles. The nanostructures simultaneously break mirror and inversion symmetries, resulting in chiral and pseudo-chiral optical responses that manifest as linear and circular polarization-dependent spectral features. Our numerical and experimental results reveal clear chiroptical effects in both near- and far-field. Near-field scanning optical microscopy confirms the excitation of polarization-selective localized plasmonic modes, with spatially distinct hot-spots lighting up under different incident polarizations. Furthermore, we demonstrate that the metasurface exhibits a measurable enantiospecific optical response when coated with thin left- or right-handed chiral overlayers. The differential circular dichroism signals observed in the presence of opposite enantiomers highlight the potential of the metasurface for label-free chiral sensing. These findings provide new insights into the interplay between structural anisotropy, pseudo-chirality, and enantioselective interactions in planar plasmonic systems. Our findings highlight the ability of planar metasurfaces to emulate chiral optical behavior without requiring volumetric 3D structures.
\end{abstract}


\justify

\section{Introduction}

Plasmonic metasurfaces have emerged as a transformative class of engineered materials capable of manipulating light at subwavelength scales, enabling unprecedented control over polarization, phase, and amplitude of electromagnetic waves.\textsuperscript{\cite{maier2007plasmonics,cai2011optical,ali2023circular,pluchery2024introduction}} Their ability to tailor optical responses through geometric design has opened new avenues in nanophotonics at both linear and nonlinear regimes.\textsuperscript{\cite{valev2016chiral,palermo2020biomolecular,valev2014nonlinear,infusino2014loss,krishna2016dye,kumar2025maximal,hentschel2017chiral}}

Among the various functionalities, the anisotropic and chiral properties of metasurfaces are of particular interest.\textsuperscript{\cite{albooyeh2023classification,palermo2024intrinsic,ali2024maximum,ali2025strong, kolkowski2021pseudochirality,jones2023dense}} Anisotropic metasurfaces allow for polarization-dependent phase shifts, facilitating applications in polarization converters, modulators, and beam steering devices.\textsuperscript{\cite{ding2021recent,lin2022high,yang2023integrated}} Chiral metasurfaces, on the other hand, exhibit asymmetric interactions with circularly polarized light, leading to phenomena such as circular dichroism and optical activity, which are crucial for chiral linear and nonlinear sensing, enantiomer discrimination, and quantum optics applications.\textsuperscript{\cite{valev2014nonlinear,rajaei2019giant,hajji2021chiral,koyroytsaltis2022detecting,garcia2019enhanced,collins2018enantiomorphing}}
Nonlinear circular dichroism based on octupolar chiral plasmonic structures has been demonstrated by way of second-harmonic generation.\textsuperscript{\cite{kolkowski2015octupolar}}  
Several studies have highlighted the emergence of pseudo-chirality  in planar metasurfaces, arising not from an intrinsic three-dimensional handedness, but from the combination of in-plane geometrical asymmetry and oblique incidence illumination. This effect, sometimes referred to as “extrinsic chirality” or “geometric pseudo-chirality,” can lead to chiroptical responses that mimic those of intrinsically chiral systems. Unlike intrinsic chirality, which is preserved regardless of the observation conditions, pseudo-chirality depends critically on the relative orientation between the structure and the incident wavevector.\textsuperscript{\cite{deng2024advances,de2015strong,plum2009metamaterials}}

The integration of anisotropic and chiral characteristics in plasmonic nanostructures can lead to devices with enhanced performance and novel functionalities.\cite{valev2013chirality,collins2017chirality} Recent advancements have demonstrated  anisotropic chiral plasmonic metamaterials capable of polarization conversion and detection, leveraging thin-film interference effects to minimize optical loss while maximizing chirality.\textsuperscript{\cite{bai2021highly,hegedus2006imaging,stamatopoulou2022reconfigurable}} These developments underscore the potential of such metasurfaces in applications ranging from biosensing to advanced imaging systems.\textsuperscript{\cite{rippa2025diagnostic, pellegrini2019superchiral}}

In recent years, there has been a growing demand for ultrasensitive platforms capable of detecting molecular chirality due to the biological relevance of enantiomers, which can exhibit drastically different pharmacological behaviors despite having identical chemical compositions.\textsuperscript{\cite{mcvicker2024chirality,lininger2023chirality, solomon2018enantiospecific,mohammadi2018nanophotonic}} Traditional techniques such as circular dichroism (CD) spectroscopy often fail to detect low concentrations of chiral molecules. To overcome this limitation, plasmonic chiral metasurfaces have been developed with enhanced chiroptical responses capable of sensing chiral molecules down to the zeptomole scale, which are orders of magnitude more sensitive than conventional CD methods.\cite{garcia2018enantiomer} For instance, Zhao et al. demonstrated twisted optical metamaterials with distinct spectral responses to opposite enantiomers, enabling unambiguous chirality detection even for molecules with inherently weak CD signals.\textsuperscript{\cite{zhao2017chirality}} Similarly, localized chiral plasmonic resonances, particularly when combined with lattice effects, have shown high enantiosensitivity, as highlighted in studies of antibody–antigen interactions. These localized modes interact strongly with the chiral environment, with the near-field distribution being highly sensitive to structural variations in the analyte, especially in the presence of birefringent chiral layers that act as sinks or sources of optical chirality. Surface morphology and defects further modulate inductive coupling between nanostructure elements, amplifying their sensing capabilities.\textsuperscript{\cite{koyroytsaltis2022detecting}} The potential of plasmonic metasurfaces has also been extended by the development of ultrathin suspended chiral bilayers, which combine high optical chirality with simplified fabrication procedures. As reported by Cen et al., these self-aligned structures achieve label-free detection of chiral molecules with unprecedented sensitivity, demonstrating that planarized metasurfaces can serve not only as high-performance circular polarizers but also as platforms for enantiomeric discrimination.\textsuperscript{\cite{cen2022ultrathin}} 

In parallel with these developments, a new trend in the field of chiral and pseudo-chiral metasurfaces has emerged, centered on the exploitation of (quasi-) bound states in the continuum (BICs and quasi-BICs).\textsuperscript{\cite{chen2023observation, zhang2022chiral, liu2019circularly, bai2025recovery,jeong2025obtuse}} BICs are non-radiating resonant states embedded within the radiation continuum, theoretically capable of confining light indefinitely. In practical systems, they manifest as quasi-BICs with exceptionally high quality (Q) factors, enabling strong field confinement alongside controlled radiation.

To fully realize the capabilities of metasurfaces in engineering and enhancing the chiroptical response, comprehensive optical characterization is essential. Far-field measurements provide insights into the overall optical response, while near-field techniques, such as scattering-based scanning near-field optical microscopy (s-SNOM), offer detailed information on local electromagnetic field distributions.\textsuperscript{\cite{bazylewski2017review,chen2015structure}} 
It is worth noting that s-SNOM measurements offer rich insight into the local electromagnetic environment of plasmonic nanostructures. However, interpreting these measurements requires careful consideration, as the detected signals often result from a combination of multiple spatially-varying near-field components rather than uniform vector fields.\textsuperscript{\cite{neuman2015mapping}} Polarization-resolved s-SNOM techniques enable the reconstruction of the local near-field vector and polarization state with nanometer resolution.\textsuperscript{\cite{schnell2010phase}} By controlling the polarization of both the illumination and detection paths, it is possible to selectively probe specific components of the sample’s permittivity tensor.\textsuperscript{\cite{kaps2023polarization,ferjani2008statistical}} The coupling efficiency between the near-field probe and mesoscopic metallic structures is highly dependent on the relative orientation of the field polarization and sample geometry.\textsuperscript{\cite{gademann2004study}} Therefore, careful consideration of polarization effects is essential for the accurate characterization of near-field distributions in plasmonic systems. These insights are particularly relevant for understanding and engineering chiral nanostructures, which rely on precise control of light–matter interactions across a broad spectral range.\textsuperscript{\cite{schnell2016real}}

In this work, we present the design, fabrication, and optical characterization of a plasmonic metasurface composed of nanostructures with the shape of fused equilateral triangles, aligned along one side, exhibiting anisotropic, chiral and pseudo-chiral optical responses. We employ both far-field and near-field (s-SNOM) techniques to investigate the optical properties under linearly and circularly polarized illumination at normal and oblique incidence, highlighting the metasurface's directional response and local field variations.
With the aid of numerical simulations, we quantify the relative contributions of chiral and anisotropic effects to the overall dichroic response.
Importantly, we also show that the metasurface enables enantiospecific detection by exhibiting distinct optical responses in the presence of left- and right-handed chiral overlayers. Despite the chiral layers having their intrinsic absorption bands outside the metasurface’s primary resonances, we observe measurable changes in the circular dichroism spectra, confirming the structure’s sensitivity to molecular handedness. These results suggest that the engineered near-field environment, with enhanced optical chirality density, plays an important role in amplifying the subtle differences between enantiomers.
Our findings contribute to the growing body of knowledge on anisotropic chiral metasurfaces and establish a pathway for their use in label-free chiral sensing, optical discrimination of enantiomers, and polarization-resolved nanophotonic applications.

\section{Results and Discussion}

\subsection{Fabrication and Far-Field Characterization under linear polarization}

Two-dimensional nanopatterns were fabricated using a Raith150 electron beam lithography (EBL) system.  The fabrication of the plasmonic metasurface followed standard EBL procedures. A 150 nm-thick layer of styrene–methyl acrylate copolymer (ZEP520A) was spin-coated onto a glass substrate pre-coated with a 15 nm indium tin oxide (ITO) layer, and subsequently baked at 170$^\circ$C for 5 minutes. Pattern definition was carried out by exposing the resist to an 18 pA electron beam with an area dose of 28 mC/cm$^2$. Development was performed by sequential immersion in methyl isobutyl ketone (MIBK) for 90 s, a 1:3 MIBK:isopropyl alcohol (IPA) solution for 60 s, and finally in pure IPA for 30 s. These steps resulted in the formation of nanostructures matching the intended design in the resist layer.
Plasmonic nanopatterns  were then obtained by electron beam evaporation of a 2 nm chromium adhesion layer followed by 55 nm of gold (SISTEC CL-400C). To remove the resist and leave only plasmonic nanostructures, a chemical lift-off process was employed by immersing the sample in acetone for 20–25 minutes, followed by a 5-minute bath in N-methyl-pyrrolidone (NMP) at 80$^\circ$C.\textsuperscript{\cite{chen2015structure}} 

\begin{figure}[!hbt]
\centering
\includegraphics[width=0.9\columnwidth]{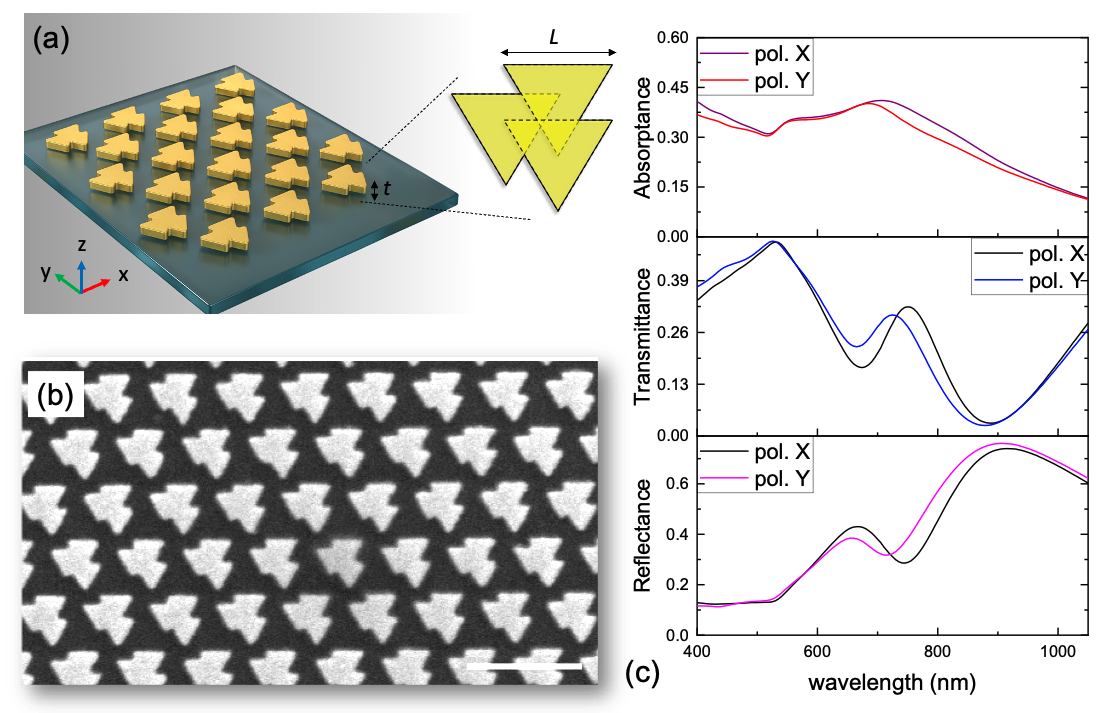}
\caption{(a) Schematic illustration of the fabricated metasurface, composed of plasmonic nanostructures periodically arranged on a planar substrate, to form a hexagonal lattice. The inset shows the 2D nanostructure geometry, defined by the lateral dimension $L$. (b) Top-view SEM image of the metasurface - scale bar: 1 $\mu$m. (c) Optical response under linearly polarized illumination along the x and y axes: absorptance (top), transmittance (middle), and reflectance (bottom) spectra, respectively.} 
\label{fig1}
\end{figure} 

Figure 1 presents the structural design and the far-field optical characterization of the engineered metasurface. The schematic in Figure 1a illustrates the periodic hexagonal arrangement of the plasmonic nanostructures deposited on a flat glass substrate. Each unit cell lacks mirror symmetry along both the x- and y-axes, introducing geometric 2D chirality and in-plane anisotropy. The presence of the glass substrate breaks the symmetry along the z-direction, rendering the metasurface effectively 3D-chiral.
The inset highlights the characteristic dimensions of the Au meta-atom, namely the lateral length of the major edge  $L$ equal to 300 nm, with overall size of approximately 450 $\times$ 430 nm, and the thickness $t$ of about 50 nm. 
The periodicity of the hexagonal lattice is 600 nm, which results in edge-to-edge separation between the neighboring nanostructures equal to approximately 150 nm and 50 - 75 nm along the x- and y-direction, respectively. These interdistances ensure a good compromise between a high surface density of the nanostructures and a spacing between the objects that avoids issues during the fabrication process. The resulting sample consists of multiple arrays, each of lateral dimensions of 200 $\mu$m $\times$ 200 $\mu$m.

The image obtained with the scanning electron microscopy (SEM - FEG 400, Philips) (Figure 1b) confirms the successful fabrication of the metasurface with high fidelity and repeatability of the nanoscale features over a large area. The non-centrosymmetric triangular elements clearly exhibit rotational asymmetry, which is expected to give rise to polarization-dependent light–matter interactions. The optical response of
the metasurface in the spectral range 400 - 1100 nm is shown in Figure 1c for normally incident light linearly polarized along the x- and y-directions. Notably, the absorptance, transmittance, and reflectance spectra are clearly polarization-dependent.
This behavior arises primarily from two key factors: the in-plane geometric anisotropy of the unit cell and the resonant plasmonic nature of the nanostructures. The elementary unit of the metasurface consists of three fused equilateral triangles arranged in a configuration that lacks both fourfold rotational symmetry and mirror symmetry along the x- and y-axes. Breaking the fourfold symmetry leads to linear anisotropy, i.e., optical response dependent on the incident linear polarization, while breaking the mirror symmetry leads to the chiroptical effects, i.e., dependence on the incident circular polarization.
At the same time, the metallic nanostructures support localized surface plasmon resonances (LSPRs); the specific triangular configuration promotes modal interference and near-field coupling among adjacent dipolar elements, enabling constructive or destructive interactions that depend sensitively on the polarization state. This leads to selective enhancement or suppression of the plasmonic modes for a given polarization, as reflected in the distinct spectral profiles of T$_x$ and T$_y$, in particular in the 650 - 850 nm range.

\subsection{Near-Field Characterization and Mode Mapping}

\begin{figure}[!hbt]
\centering
\includegraphics[width=1\columnwidth]{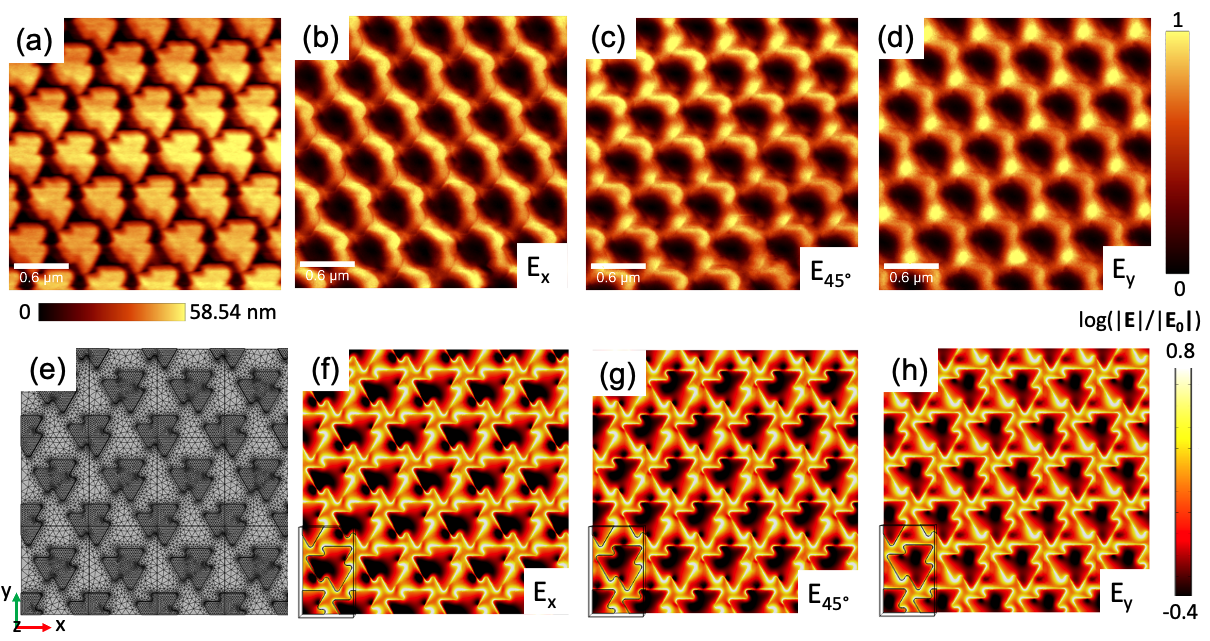}
\caption{Near-field optical characterization and numerical simulations of the metasurface. (a) Atomic force microscopy (AFM) topography of the metasurface. (b–d) Near-field amplitude maps acquired via scattering-type scanning near-field optical microscopy (s-SNOM) under linearly polarized illumination along the $x$, $45^\circ$, and $y$ directions, respectively. (e) Metasurface geometry used in the numerical model. (f–h) Simulated electric field intensity distributions  at the surface of the nanostructures for incident polarizations along the $x$, $45°$, and $y$ directions, respectively.}
\label{fig2}
\end{figure}

To gain insight into the local electromagnetic behavior of the metasurface, we performed near-field optical measurements using scattering-type scanning near-field optical microscopy (s-SNOM - alpha300 by WITec - $\lambda_{exc}$= 532 nm). Figures \ref{fig2}b-d show amplitude maps acquired under illumination by normally-incident light, linearly polarized along three distinct directions: $x$ ($0^\circ$), $45^\circ$, and $y$ ($90^\circ$). The corresponding topography, shown in Figure \ref{fig2}a, obtained by atomic force microscopy (AFM), serves as a reference for correlating the optical features with the geometry of the nanostructures. 
The near-field maps reveal a remarkably high density of localized hot-spots surrounding the nanostructures, particularly at their tips and edges. These regions of intense electromagnetic field enhancement are indicative of strong plasmonic resonances and are important for enhancing light–matter interaction processes, for example in sensing and nonlinear optical applications. Strikingly, the spatial distribution and intensity of the hot-spots vary significantly with the incident polarization. For instance, illumination along $x$ tends to activate hot-spots preferentially at the left and right tips of the triangular structures, whereas illumination along $y$ enhances the response at the top and bottom edges. Intermediate behavior is observed for the $45^\circ$ case, where a more symmetric excitation is evident.

To support and interpret the experimental results, full-wave numerical simulations were carried out using the Finite Element Method (FEM) in COMSOL Multiphysics (Figures \ref{fig2}f-h). A 3D model was built to accurately reproduce the geometry of the metasurface, including the gold nanostructures, the glass substrate (modeled as a semi-infinite medium), and any overlying dielectric or chiral layers.
The material properties of gold were defined using a wavelength-dependent complex permittivity interpolated from experimental data (Johnson and Christy),\textsuperscript{\cite{johnson1972optical}} while the refractive indices of the surrounding media were modeled as either constant or dispersive, depending on the specific configuration.
A linearly or circularly polarized plane wave was used as the excitation source, impinging at either normal or oblique incidence, depending on the case. Perfectly matched layers (PMLs) were applied on the top and bottom of the simulation domain to avoid spurious reflections.

To obtain the electric and magnetic field distributions, a frequency-domain study was performed across the desired spectral range. 

Additional details regarding the implemented model are reported in the Supporting Information (Figure S1). The model, shown in Figure \ref{fig2}e,  reproduces the metasurface geometry, and the calculated near-field distributions match the experimental patterns with high qualitative agreement. The simulated field maps, plotted as $\log(|\mathbf{E}|/|\mathbf{E_0}|)$ (where $|\mathbf{E_0}|$ is the incident field amplitude), confirm the polarization-sensitive hot-spot distribution and further highlight the intense field confinement achievable in specific regions of the nanostructures. Profiles corresponding to the SNOM and AFM measurements are included in the Supporting Information (Figure S2).
This variation in the near-field maps confirms that the metasurface not only breaks spatial symmetry but also couples the incident light selectively to different localized plasmonic modes, generating field distributions that strongly depend on the incident linear polarization. This observation is consistent with expectations for planar asymmetric structures, where broken transverse symmetry enables polarization-dependent coupling between the incident light and plasmonic modes.

\subsection{Structural chiral and pseudo-chiral response under CPL illumination}

To further unravel the complex polarization-dependent response of the metasurface, we investigated its behavior under circularly polarized light (CPL), for both co- and cross-polarized configurations, at various angles of incidence. This approach enables the decoupling of the chiral and anisotropic contributions to the circular dichroism (CD), and reveals the pseudo-chiral behavior that emerges in planar structures under tilted excitation.\textsuperscript{\cite{plum2009extrinsic,valev2013chirality, bertolotti2015second}}

\begin{figure}[!hbt]
\centering
\includegraphics[width=1\columnwidth]{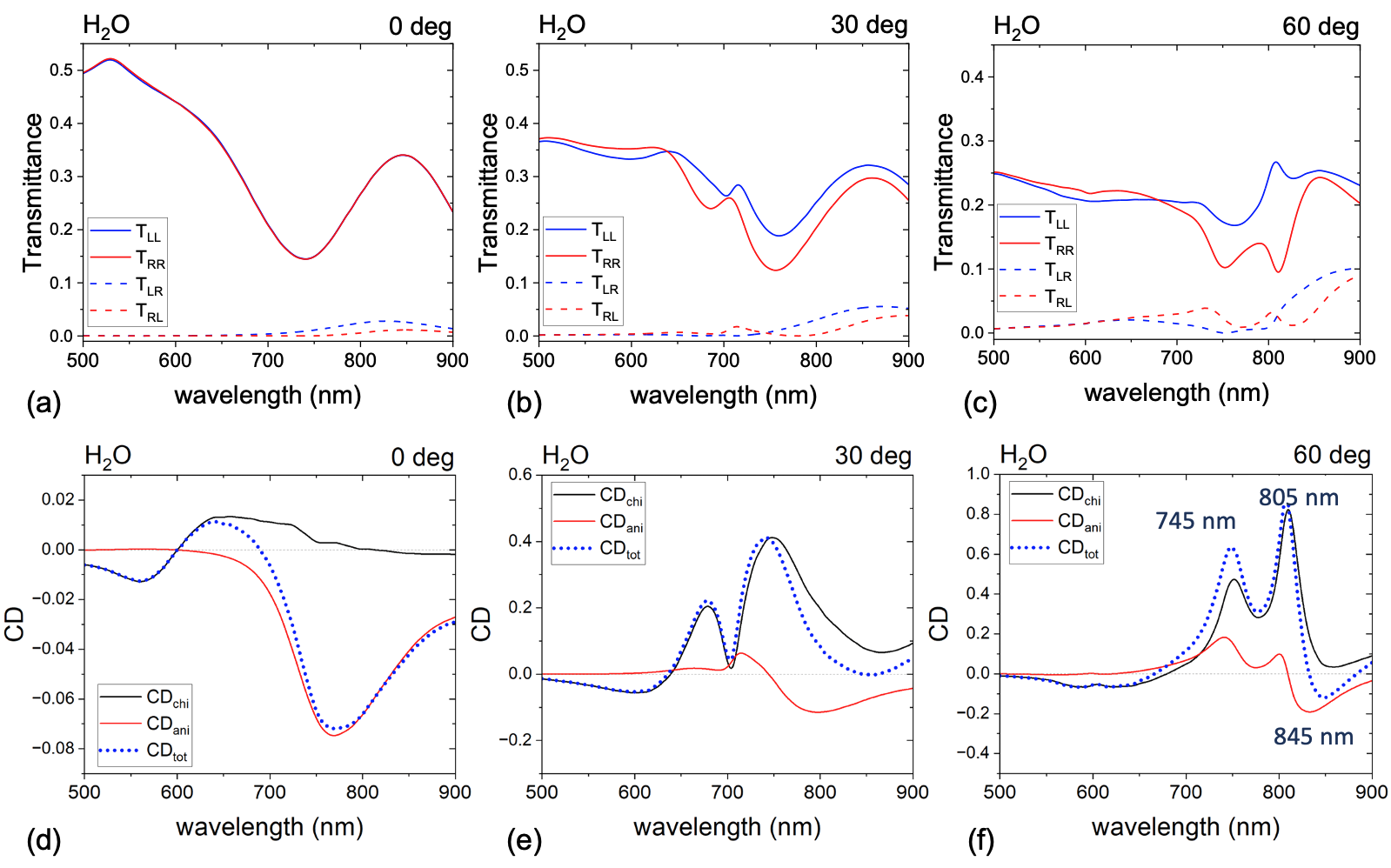}
\caption{Far-field transmittance and CD response of the metasurface immersed in water under circularly polarized illumination. 
(a–c) Transmittance spectra for co- and cross-polarized CPL components (\(t_{\mathrm{LL}}\), \(t_{\mathrm{RR}}\), \(t_{\mathrm{LR}}\), and \(t_{\mathrm{RL}}\)) at incidence angles of 0°, 30°, and 60°, respectively. 
(d–f) Decomposition of the total circular dichroism (\(CD_\mathrm{tot}\), blue dashed) into the chiral contribution (\(CD_\mathrm{chi}\), black) and the anisotropic contribution (\(CD_\mathrm{ani}\), red), for the same three incidence angles.}
\label{fig3}
\end{figure}

Figure \ref{fig3}a-c reports the transmission spectra for left- and right-CPL, resolved into the co-polarized ($t_{LL}, t_{RR}$) and cross-polarized ($T_{LR}, T_{RL}$) components, at 0$^\circ$, 30$^\circ$, and 60$^\circ$ incidence angles. The data presented in the main text refer to water as the surrounding medium. Data corresponding to the metasurface in air are provided in the Supporting Information (Figure S3).

To quantify the contributions of different mechanisms to the observed chiroptical response, we calculated the total CD spectrum as the sum of the chiral component (CD$_\text{chi}$), associated with the structural chirality, and anisotropic component (CD$_\text{ani}$), arising from the in-plane birefringence of the nanostructures. These quantities were evaluated by using the following expressions:\textsuperscript{\cite{albooyeh2023classification}}



\begin{equation}
CD_{chi} = \frac{T_{LL}-T_{RR}}{\frac{1}{2}(T_{LL}+T_{RL}+T_{LR}+T_{RR})} \label{eq2}
\end{equation}

\begin{equation}
CD_{ani} = \frac{T_{RL}-T_{LR}}{\frac{1}{2}(T_{LL}+T_{RL}+T_{LR}+T_{RR})} \label{eq1}
\end{equation}

The total CD (CD$_{tot}$) can be calculated as the sum of these two contributions as: CD$_{tot}$=CD$_{ani}$+CD$_{chi}$. 
The result shown in Figure \ref{fig3}d demonstrate that even at normal incidence, a weak dichroic response is present, primarily due to the anisotropic contribution CD$_\text{ani}$. However, as the angle of incidence increases (Figures \ref{fig3}e-f), both CD$_\text{chi}$ and CD$_\text{ani}$ exhibit a dramatic enhancement. At 60$^\circ$, the total CD reaches values close to unity in specific spectral regions, clearly indicating the emergence of pseudo-chiral effects, which are symmetry-forbidden at normal incidence but activated at oblique angles due to symmetry breaking in the incident wavevector.
This analysis confirms that the metasurface supports both chiral and anisotropic optical modes, whose spectral signatures and intensities can be modulated by tuning the incidence angle. 

 Control simulations performed for a symmetric configuration, where the metasurface is surrounded by air on both sides at normal incidence, show a complete suppression of the chiral CD component (CD$_{chi}$), and the total CD is dominated exclusively by the anisotropic term - see Figure S4 in the Supporting Information. In the absence of the substrate, the system retains the in-plane birefringence but lacks the necessary vertical asymmetry to activate chiral near-field modes. 
 The substrate breaks mirror symmetry along the propagation direction, thereby enabling the coupling of circularly polarized light to chiral plasmonic resonances. It also modifies the local boundary conditions, enhancing field confinement and boosting the optical chirality density near the metal–dielectric interface. These findings underscore that the substrate is not merely a passive support but plays an active and essential role in unlocking the chiroptical functionality of planar metasurfaces.\textsuperscript{\cite{gorkunov2025substrate}}

\begin{figure}[bt]
\centering
\includegraphics[width=1\columnwidth]{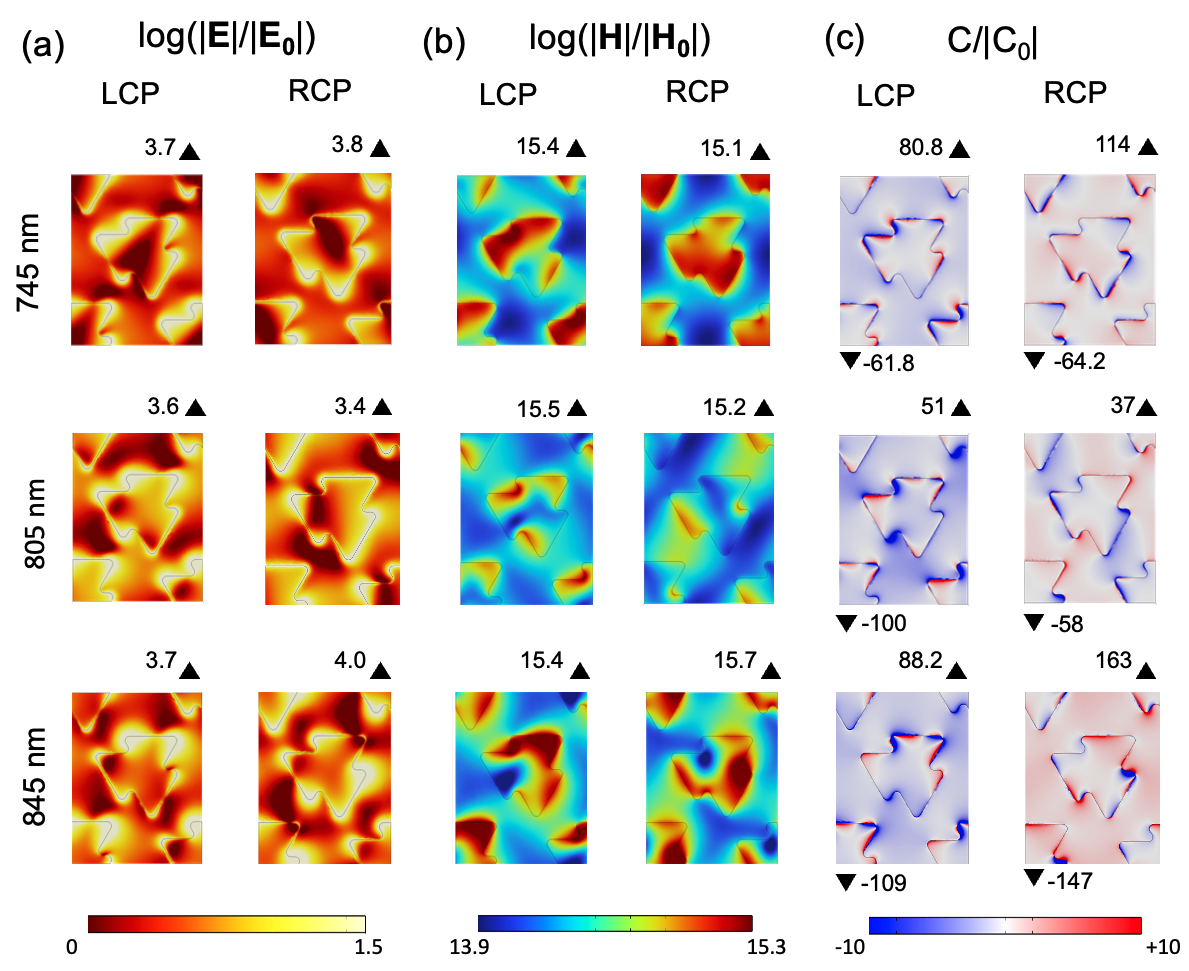}
\caption{Simulated near-field distributions under circularly polarized illumination.
Electric field intensity $\log|\mathbf{E}|/|\mathbf{E}_0|$ (a), magnetic field intensity $\log|\mathbf{H}|/|\mathbf{H}_0|$ (b), and normalized optical chirality density $C/|C_0|$ (c), evaluated at three wavelengths: 745 nm, 805 nm, and 845 nm.  The maximum and minimum values are reported above and below each map.}
\label{fig4}
\end{figure}

Figure \ref{fig4} shows the near-field distributions and optical chirality maps computed using  COMSOL Multiphysics.
The optical chirality density 
($C$) was computed from the simulated fields using the following expression: $C=-\frac{\epsilon_0 \omega}{2}Im(\textbf{E*}\cdot\textbf{H})$, where $\epsilon_0$ and $\omega$ are the vacuum permittivity and the angular frequency, respectively; 
$\textbf{E}$ and $\textbf{H}$ denote the complex electric and magnetic field vectors.\textsuperscript{\cite{tang2010optical,tang2011enhanced,solomon2020nanophotonic}}

To better understand the physical origin of the chiroptical response observed in the far-field, we analyzed the spatial distribution of the electromagnetic fields and optical chirality density at selected resonance wavelengths. This analysis is related to the optical response at the incidence angle of 60°. Figure \ref{fig4} shows simulated maps of (a) the electric field intensity, (b) the magnetic field intensity, and (c) the normalized optical chirality density, on a horizontal plane located above the top surface of the nanostructures, for both left- and right-circularly polarized (LCP and RCP) illumination at 745 nm, 805 nm, and 845 nm. These three wavelengths were selected based on the characteristic points of the CD spectrum presented in Figure \ref{fig3}f.
The electric field maps (Figure \ref{fig4}a) reveal a pronounced localization of near-field hot-spots around the nanostructures, especially near the sharp corners and inner notches. These hot-spots reach intensity enhancements exceeding an order of magnitude with respect to the incident field, and exhibit distinct spatial distributions depending on the handedness of the incident circular polarization. Similar features are observed in the magnetic field maps (Figure \ref{fig4}b), although with different spatial patterns and symmetries, contributing to the near-field asymmetry required for optical chirality enhancement. Field intensities were normalized to the incident field magnitudes and plotted in logarithmic scale to clearly reveal the hot-spot distribution.

Most importantly, the calculated optical chirality density, (Figure \ref{fig4}c) demonstrates a substantial local amplification near the nanostructures, reaching values over two order of magnitude larger than the chirality of a circularly polarized plane wave in free space, $C_0=\pm E_0^2 \epsilon_0 \omega/(2c)$. A value of $|C|/|C_0|$ exceeding 1 can be obtained by exploiting the near-field enhancement by plasmonic nanostructures.\textsuperscript{\cite{kelly2018controlling,tabouillot2022near,gilroy2022superchiral}}     
Similarly to the local electric and magnetic fields, the spatial distribution of $C$ is highly asymmetric and strongly dependent on the handedness of incident circularly-polarized light. Notably, regions of enhanced optical chirality (superchirality) are tightly confined to the edges of the nanostructures, suggesting the presence of chiral plasmonic resonances enabled by the interplay of electric and magnetic near-fields.
To quantitatively assess the degree of optical chirality enhancement, we computed the maximum normalized chirality density $C/|C_0|$. As shown in Figure \ref{fig4}c, at the resonance wavelength of 845 nm, the simulated values of $C/|C_0|$ reach peaks exceeding 150, with localized regions consistently above 100 across multiple structural hot-spots. These values confirm the generation of superchiral fields with intensities more than two orders of magnitude higher than those of CPL in homogeneous media. Importantly, this amplification is not spatially uniform but concentrated at the nanoscale edges and gaps of the plasmonic motifs, particularly where electric and magnetic fields overlap and contribute constructively to $C$. 

The formation of electromagnetic hot-spots within the analyzed metasurface arises from the interplay between the plasmonic geometry of the nanostructures, the material properties, and the excitation conditions. A key mechanism behind this phenomenon is the excitation of LSPRs, which occur when the frequency of the incident light matches the natural oscillation frequency of free electrons on the surface of the metal nanostructure. This resonance leads to a strong concentration of the electric field near the metal surface. Additionally, the so-called lightning rod effect, resulting from the accumulation of surface charges at locations with high surface curvature, such as sharp corners and recessed edges, leads to an additional local field enhancement. Beyond the individual nanostructure effects, the periodic arrangement and dense packing fosters near-field coupling between the adjacent elements. 
The distribution and intensity of these field enhancements are highly dependent on the wavelength and polarization of the incoming light, and are especially sensitive to the symmetry and orientation of the incident wave.
Under circularly polarized illumination, the asymmetric geometry of the metasurface results in a distinct near-field response for left- and right-circularly polarized light (LCP and RCP). The lack of mirror symmetry causes a differential coupling of the two polarizations to the plasmonic modes, producing polarization-dependent field distributions. This asymmetry becomes more pronounced at oblique incidence, where the in-plane component of the wavevector breaks additional symmetries, activating the so-called pseudo-chiral modes that are not excited under normal incidence.
One of the most critical consequences of this near-field localization is its impact on the optical chirality density, a quantity that depends on the spatial and phase overlap between the electric and magnetic fields. Specifically, high optical chirality emerges in regions where both the electric and magnetic field amplitudes are simultaneously enhanced and non-orthogonal. These regions, where the product $\text{Im}(\mathbf{E}^* \cdot \mathbf{B})$ is maximized, define the so-called \textit{chiral hot-spots}, which are typically concentrated near the edges and vertices of the nanostructures. The simulated optical chirality maps (Figure \ref{fig4}c) confirm this behavior, revealing that the metasurface generates strong superchiral near-fields, often exceeding the common value obtained with circularly polarized plane waves by more than two order of magnitude.

\subsection{Quantifying Enantiomeric Contrast through Differential Chiroptical Response}

To explore the enantioselective potential of the metasurface, we numerically investigated how the total chiroptical response of the metasurface is influenced by the presence of a chiral layer in its close proximity. The angle of incidence considered in the analysis is 60$^\circ$, while the azimuthal angle of incidence, (i.e., the angle between the x-axis and the wavevector projection on the xy-plane), is set to zero. 
Specifically, we simulated the system's anisotropic CD ($\mathrm{CD}_{\mathrm{ani}}$), chiral CD ($\mathrm{CD}_{\mathrm{chi}}$), and total CD ($\mathrm{CD}_{\mathrm{tot}}$) in three configurations: with a non-chiral dielectric overlayer, and with left-handed (LH) and right-handed (RH) chiral layers placed above the nanostructured surface. In all three cases, the overlayer was modeled as a 70 nm thick layer. Details regarding the modeling of the chiral layers are provided in the Supporting Information (Figure S5).

\begin{figure}[!hbt]
\centering
\includegraphics[width=1\columnwidth]{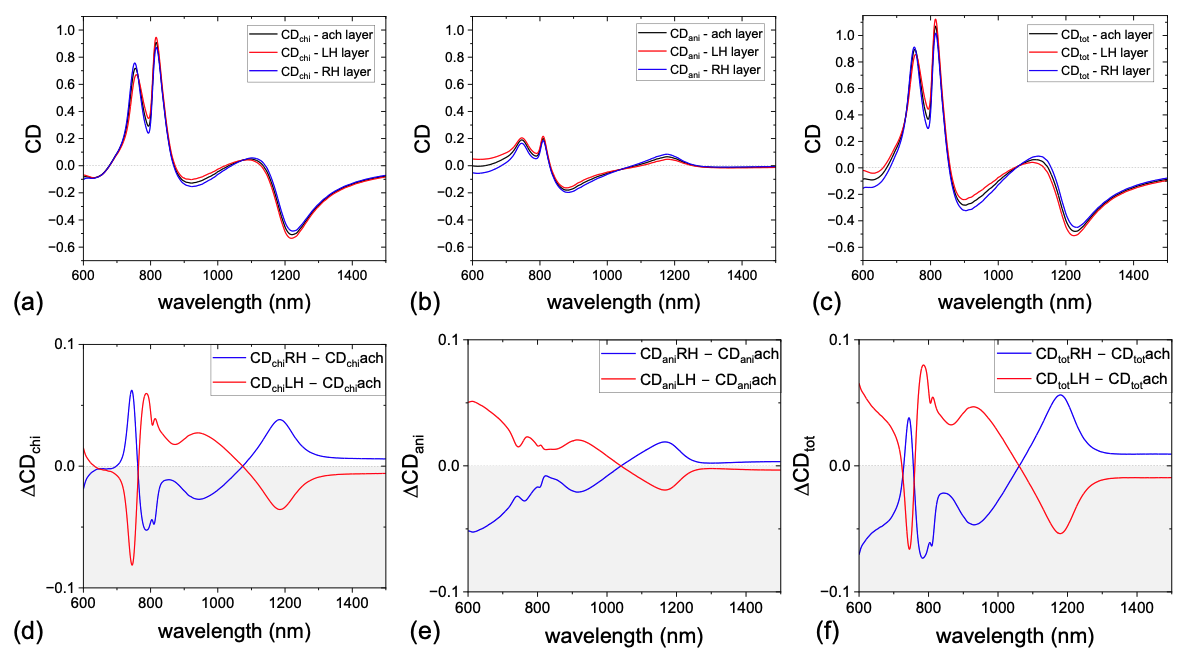}
\caption{Circular dichroism (CD) of the metasurface with different overlayers: achiral, left-handed (LH), and right-handed (RH). Panels (a–c) show, respectively, the chiral (\textit{CD\textsubscript{chi}}), anisotropic (\textit{CD\textsubscript{ani}}), and total (\textit{CD\textsubscript{tot}}) CD spectra for the three overlayer conditions. Panels (d–f) display the corresponding differential CD signals: RH–achiral (blue), and LH–achiral (red), highlighting the distinct spectral signatures of the enantiomers.}
\label{fig5}
\end{figure}

The achiral layer, used as a reference, was modeled as a racemic mixture, i.e., an optically inactive material with no net chirality. It shares the same dispersion and refractive index properties as the chiral layers, allowing for a consistent comparison.
To simulate the optical response of a thin chiral layer deposited on the metasurface, we adopted refractive index values of $n$=1.3571 for the real part and $k$=0.02 for the imaginary part. These values are consistent with those reported for protein-based materials and organic chiral films in the visible spectral range, particularly when the thickness is below 100 nm.\textsuperscript{\cite{haynes2016crc}} The chosen parameters account for a weak but non-negligible absorption, as typically observed in biological materials under visible illumination. Furthermore, the chiral layer was modeled with a non-zero chirality parameter $\xi$, peaking at 450 nm, to mimic the dispersive behavior of naturally occurring chiral media. This configuration allows us to assess the impact of molecular chirality on the optical response of the metasurface, while keeping the dielectric background consistent with realistic biological conditions.

 As shown in Figures \ref{fig5}a–c, all three CD components exhibit clear spectral modulations due to the presence of chiral material, with the RH and LH overlayers producing measurably different responses.

The anisotropic CD exhibits slight spectral shifts and amplitude changes depending on the handedness of the superstrate.
The most pronounced modulations appear in $\mathrm{CD}_{\mathrm{chi}}$, which reflects the interaction between CPL and the broken-symmetry surface in a chiral environment.
The total CD, combining both contributions, shows the most significant differences between the three configurations. The introduction of a LH or RH layer results in strong differential modulation, especially near the spectral regions associated with plasmonic resonances.  Importantly, the spectral signatures in the LH and RH cases are different, suggesting that the metasurface can discriminate between enantiomers based on their optical activity.
To isolate these differences, we computed differential CD signals ($\Delta$CD) by subtracting pairwise spectra (RH–achiral, LH–achiral), as plotted in Figures \ref{fig5} d–f.

The results, shown in Figures \ref{fig5} d-f, reveal that the metasurface is sensitive to the handedness of the overlaid chiral medium.
From the comparison between RH–achiral and LH–achiral spectra, we can isolate the specific contribution of the chiral overlayer to the overall CD response. This differential approach highlights how the presence of a chiral layer modulates the CD signal in a handedness-specific way, with respect to the racemic one. The difference becomes even more evident in the total CD (Figure \ref{fig5}f), which is predominantly influenced by the $\mathrm{CD}_{\mathrm{chi}}$, with only a minor contribution from the anisotropic part.

Interestingly, although the chiral overlayer exhibits a characteristic dichroic band centered around 450 nm, well outside the primary resonance range of the plasmonic metasurface, it can still influence the optical response of the system at longer wavelengths.\textsuperscript{\cite{govorov2012theory}} This is because the presence of a chiral material modifies the local electromagnetic boundary conditions and alters the near-field distribution of the metasurface, even in spectral regions where the overlying layer itself is nominally non-resonant. In particular, plasmonic resonances are highly sensitive to the local refractive index environment. The chirality of the overlayer introduces a handedness-dependent refractive index that subtly shifts the phase and amplitude of the scattered fields. 

\begin{figure}[!hbt]
\centering
\includegraphics[width=1\columnwidth]{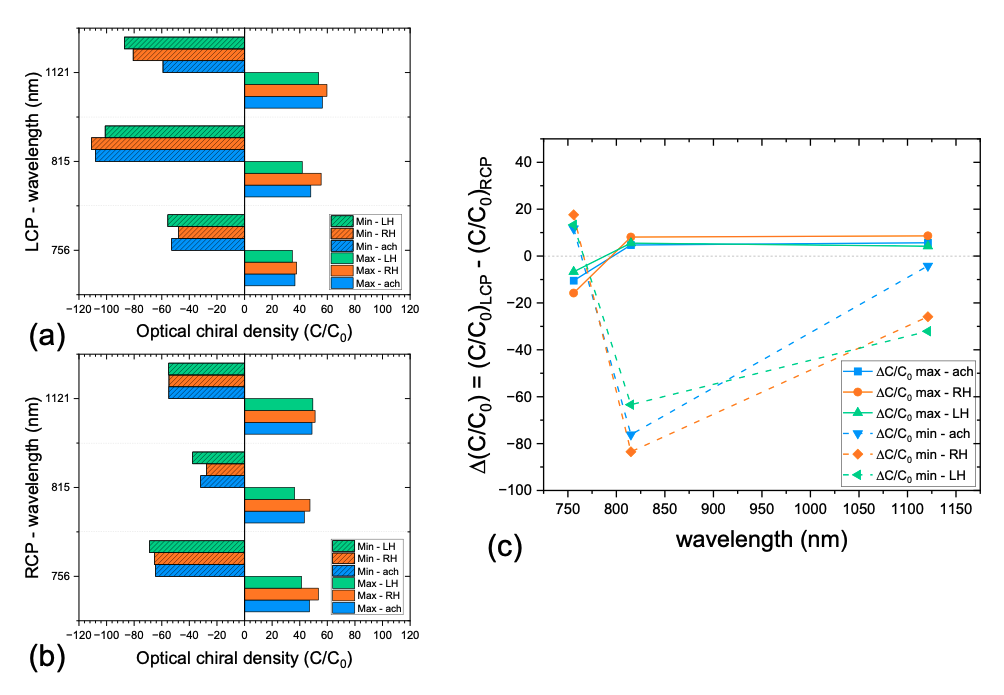}
\caption{Optical chirality density above the metasurface. Maximum (solid) and minimum (patterned) values of the normalized chirality density $C/C_0$ under (a) LCP and (b) RCP illumination for the  achiral, RH, and LH overlayers at three different wavelengths. (c) Differential chirality $\Delta(C/C_0)=(C/C_0)_{LCP}-(C/C_0)_{RCP}$ calculated for the maxima (solid lines) and minima (dashed lines).}
\label{fig6}
\end{figure}
To gain deeper insight into the enantioselective nature of the near-field response, we performed a quantitative analysis of the optical chirality density 
$C$, normalized to its free-space value $C_0$, extracted at a fixed height above the nanostructures and at three specific wavelengths (Figure S6). For each excitation polarization (LCP and RCP), we computed the maximum and minimum values of $C$ for the three distinct configurations considered (achiral, RH, and LH layers), see Figure \ref{fig6}.
Under LCP illumination (Figure \ref{fig6}a) the RH layer consistently yields the highest $C/C_0$ maxima across all the wavelengths, reaching up to $\sim$60 at 1121 nm. In contrast, the LH layer exhibits slightly suppressed maxima, indicating a reduced field overlap with the handedness of the incident wave. The difference between the RH and LH cases is most pronounced at 815 nm, reaching 13.7. A similar trend is observed in the minima of $C/C_0$, which become more negative in the RH configuration (e.g., –111 vs. –101 at 815 nm), confirming a stronger superchiral response.
Under RCP illumination (Figure \ref{fig6}b) the trend reverses in a consistent and enantioselective manner. Here, the LH layer yields the lowest minima (e.g., –68.9 at 756 nm), while the RH case shows reduced negative enhancement. The maxima similarly reflect this handedness preference, with the RH layer producing higher $C/C_0$ values than the LH at all the wavelengths.
To explicitly capture this handedness dependence, we calculated the differential chirality density  $\Delta(C/C_0)=(C/C_0)_{LCP}-(C/C_0)_{RCP}$ for both maxima and minima (Figure \ref{fig6}c). The RH layer shows the largest positive $\Delta(C/C_0)_{max}$ and the most negative $\Delta(C/C_0)_{min}$, confirming a stronger differential response between LCP and RCP excitation. This asymmetry is particularly evident at 815 nm, where $\Delta(C/C_0)_{min}$ reaches –85 for the RH case, compared to –76 and –63 for the achiral and LH layers, respectively.
These results demonstrate that the metasurface–overlayer system supports highly tunable enantioselective near-fields, with a differential optical chirality response that is both polarization- and material-dependent. Such selective amplification of $C$ near the surface under specific handedness conditions provides a robust metric for quantifying enantiomeric contrast and supports the platform’s potential for label-free, wavelength-resolved chiral sensing.

The abova near-field amplification mechanism underlies the ability of metasurfaces to resolve weak chiral signals. This approach resonates with recent strategies in the literature that leverage the near-field enhancement and symmetry properties of chiral metasurfaces. For example, Zhao et al.,\textsuperscript{\cite{zhao2017chirality}} demonstrated that tailored near-fields in chiral metamaterials can amplify chiral–chiral interactions with analytes, producing molecular CD responses two orders of magnitude stronger than those detected by conventional CD spectroscopy. Their method also enables background-free detection of molecular chirality by using pairs of metamaterials with opposite handedness. Similarly, Cen et al.,\textsuperscript{\cite{cen2022ultrathin}} proposed a differential CD analysis to suppress unwanted signals from the metasurface itself and isolate the response of the chiral molecules. In both cases, the differential strategy proves critical for enhancing the detection sensitivity and unambiguously identifying the sign of the chiral response. Our findings align with these reported studies, demostrating the ability to extract such subtle differential chiroptical responses, which confirms that the metasurface design is highly suitable for enantioselective sensing. However, our current investigation employs a single metasurface configuration. To further increase the specificity and robustness of the chiral detection, we plan to extend this approach by testing the mirror-image counterpart of the metasurface. This will allow us to perform a full differential analysis between enantiomorphic structures and further enhance the platform’s capacity for background-free, label-free chiral sensing. Overall, this platform holds promise for applications in surface-enhanced circular dichroism spectroscopy and ultrasensitive biomolecular analysis at low concentrations.

\section{Conclusion}

In summary, we have demonstrated that a planar plasmonic metasurface combining structural chirality and in-plane anisotropy exhibits a pronounced polarization-dependent response, which is further enhanced at oblique incidence due to the pseudo-chiral effects, with clear manifestations in both the near-field optical intensity and the far-field scattering properties. Full-wave simulations and near-field SNOM measurements reveal dense networks of electric and magnetic hot-spots whose spatial overlap gives rise to local optical chirality densities exceeding those of free-space CPL by more than two orders of magnitude. 
By further showing that the metasurface’s chiroptical response can be modulated in a handedness-specific manner through the addition of thin chiral overlayers, we establish its potential as a label-free platform for enantioselective sensing and spectroscopy. 

Looking ahead, our current efforts are focused on extending the metasurface design toward coherent nonlinear chiroptical spectroscopy. Ongoing experiments based on second-harmonic generation (SHG) are being conducted to probe the enantiosensitive near-field response under circularly polarized excitation. These studies aim to quantify the enhancement of nonlinear chiral contrast at specific hot-spots and assess the potential for multiplexed detection of chiral analytes at the nanoscale.

Additionally, fabrication of analogous metasurfaces operating in the ultraviolet and mid-infrared regimes is currently underway, targeting application-relevant spectral windows for biomolecular fingerprinting and vibrational chiral sensing. These developments will be instrumental for translating our platform into practical tools for chemical analysis, diagnostics, and enantioselective catalysis.



\medskip
\textbf{Supporting Information} \par 
Supporting Information is available from the Wiley Online Library or from the author.

\medskip
\textbf{Acknowledgements} \par 
The authors are grateful to Dr. Giovanni Desiderio for SEM measurements and analysis. 
A.G. acknowledges financial support from the “NLHT- Nanoscience Laboratory for Human Technologies” - (POR Calabria FESR-FSE 14/20 – CUP: J22C14000230007) and POS RADIOAMICA project funded by the Italian Minister of Health (CUP: H53C22000650006). D.M.A. acknowledges financial support from the project "PRIN - 2022P9F79R" (CUP: H53D23000830006). The authors gratefully acknowledge the support for this work from European Union - NextGenerationEU Call PRIN2022 Development of a plasmonic nanobiosensor for the rapid diagnosis of Shiga toxin producing E. coli human infections at the point of care - SENSOSTEC (CUP B53D23019990006).

\medskip
\textbf{Conflict of interest} \par 
The authors declare no conflict of interest.
\medskip

%
\bibliographystyle{MSP}
\bibliography{sample}



\end{document}